\documentclass[twocolumn,amsmath,amssymb]{revtex4}
\usepackage{graphicx}
\usepackage{bm}
\usepackage{amsmath,amssymb} 

\begin{document}

\title{Spectroscopy of vibrational modes in metal nanoshells} 

\author{Arman S. Kirakosyan$^{a,b}$, Tigran V. Shahbazyan$^a$}

\affiliation{$^a$Department of Physics, Jackson State University, Jackson, MS
39217, USA\\
$^b$Department of Physics, Yerevan State University, 1 Alex Manoogian
St., Yerevan, 375025, Armenia}  

%
\begin{abstract}
We study the spectrum of vibrational modes in metal
nanoparticles with a dielectric core. Vibrational modes are excited
by the rapid heating of the particle lattice that takes place after
laser excitation, and can be monitored by means of pump-probe
spectroscopy as coherent oscillations of transient optical spectra. In
nanoshells, the presence of two metal surfaces 
results in a substantially different energy spectrum of acoustic
vibrations than for solid particles. 
We calculated the energy spectrum as well as the damping of nanoshell
vibrational modes.
The oscillator strength of fundamental breathing mode is larger than
that in solid nanoparticles. At the same time, in very thin
nanoshells, the fundamental mode is overdamped due to instantaneous
energy transfer to the surrounding medium.
\end{abstract}
\maketitle

\section{Introduction}
\label{intro}

Acoustic vibrational modes in nanoparticles are impulsively excited by
a rapid heating of the lattice that takes place after laser
excitation\cite{vallee-jcp99,hartland-jcp99,vallee-jcp01,elsayed-nl04,hartland-bimetal1,hartland-bimetal2}. 
After initial period of rapid expansion, a nanoparticle  
undergoes radial contractions and expansions around the new equilibrium.
The periodic change in nanoparticle volume translates into a modulation
in time of the surface plasmon resonance (SPR) energy that dominates
nanoparticle optical absorption spectrum. The spectrum of vibrational
modes manifests itself via coherent oscillations of differential
transmission at SPR energy measured using ultrafast pump-probe
spectroscopy\cite{vallee-jpcb01}. Since the size of
laser spot is usually much larger than 
nanoparticle diameter, the initial expansion is homogeneous so that
predominantly the fundamental ($n=0$) breathing mode, corresponding to
oscillations of nanoparticle volume as a whole, is excited. The lowest
excited ($n=1$) mode has weaker oscillator strength ($\approx 1/4$ of
that for $n=0$) and thus is more difficult to observe. When
nanoparticle is embedded in a dielectric medium, the oscillations are
damped due to the transfer of latice energy to acoustic waves in
surrounding dielectric. In solid and bimetallic particles, the size
dependences of eigenmodes energy and decay rate are similar -- both
are inversely proportional to nanoparticle radius\cite{dubrovskiy81}.

Here we study the vibrartional modes of metal nanoshells. These recently
manufactured metal particles with dielectric
core\cite{halas-prl97} attracted much interest due 
to unique tunability of their optical properties. By varying the shell
thickness during the manufacturing process, the SPR can be
tuned in a wide energy interval \cite{aden51}. Recent pump-probe
measurements of vibrational modes dynamics in gold nanoshells submerged in
water revealed characteristic oscillation pattern of differential
transmission. However, the oscillations period and amplitude as well 
as their damping were significantly larger than those for solid
nanoparticles. We perform detailed analysis of energy spectrum of
lowest vibrational modes of a nanoshell in a dielectric medium. We
find that the modes eigenenergies exhibit a strong dependence on
nanoshell aspect ratio, $\kappa=R_1/R_2$, where $R_1$ and $R_2$ are
inner and outer radii, respectively. Specifically, for thin
nanoshells, the fundamental mode energy is consirerably lower than for
solid particles while the damping is significantly larger. At the same
time, in the thin shell limit, the fundamental mode carries the 
{\em entire} oscillator strength  which results in an enhanced
oscillations amplitude as compared to solid particles. The analysis
also reveals two regimes, where the spectrum is dominated by nanoslell
geometry or by surrounding medium, with a sharp crossover governed by
the interplay between aspect ratio and impendance.

\section{Spectrum of vibrational modes of a spherical shell}
\label{sec:spectrum}

We consider radial normal modes of a spherical nanoshell with dielectric
core extending up to inner radius $R_1$ in a dielectric medium beyond
outer radius $R_2$. The acoustical properties of the system are
characterized by the densities $\rho^{(i)}$ and the longitudinal and
tranverse sound velocities $c_{L,T}^{(i)}$, where $i=c,s,m$ stands for
core, shell, amd medium, respectively. The radial displacement 
$u(r)$ is determined from Helmholtz equation (at zero angular
momentum) \cite{L&L-elast} 
\begin{equation}
u'' + \frac{2u'}{r} + k^2u = 0,
\label{Helm}
\end{equation}
where $k = \omega/c_L$ is the wave-vector, with the boundary conditions
that the displacement $u$ and the radial component of stress tensor,
\begin{equation}
\sigma = \rho \Biggl[c_L^2 u' + (c_L^2 - 2 c_T^2)\, \frac{2u}{r}
\Biggr],
\label{stress}
\end{equation}
are continuous at the core/shell and shell/medium interfaces.
In the three regions divided by shell boundaries, the displacement has
the form  
\begin{eqnarray}
\label{regions}
&&
u^{(c)} \sim \frac{\partial}{\partial r} \frac{\sin k^{(c)}r}{r},
~~~
u^{(s)} \sim \frac{\partial}{\partial r} \frac{\sin (k^{(s)}r+\phi)}{r},
~~~
\nonumber\\
&&
u^{(m)} \sim \frac{\partial}{\partial r} \frac{e^{ik^{(m)}r}}{r},
\end{eqnarray}
where $\phi$ is the phase mismatch. The corresponding eigenenergies
are, in general, complex due to energy transfer to the outgoing wave in
the surrounding medium. After matching $u(r)$ and $\sigma(r)$ at $r=R_1,
R_2$, we obtain the following equations for eigenvalues $\xi=kR_2$

\begin{eqnarray}
\hspace{-1mm}
\frac{\xi^2 \kappa^2}{\xi \kappa \cot(\xi \kappa + \varphi) - 1}
 - 
\frac{\eta_{c}\xi^2 \kappa^2}{(\xi \kappa/\alpha_{c}) \cot(\xi
  \kappa/\alpha_c) - 1} +\chi_c =0,
\nonumber \\ 
\frac{\xi^2}{\xi \cot(\xi + \varphi) - 1} 
 + 
\frac{\eta_{m}\xi^2}{1 + i \xi/\alpha_m} + \chi_m = 0, 
\qquad
\label{breathmodes}
\end{eqnarray}
where $\kappa  =  R_1/R_2$ is nanoshell aspect ratio, and the
parameters 
\begin{eqnarray}
\alpha_{i} = c_{L}^{(i)}/c_{L}^{(s)}, 
~~
\eta_{i} = \rho^{(i)}/\rho^{(s)}, 
~~
\chi_{i} = 4 (\beta_s^2- \eta_i \delta_i^2)
\nonumber \\
~~
\beta_{i}=c_{T}^{(i)}/c_{L}^{(i)},
~~
\delta_{i}=c_{T}^{(i)}/c_{L}^{(s)}.
\qquad \qquad \qquad \qquad 
\label{parameters}
\end{eqnarray}
characterize the metal/dielectric interfaces. 
In the ideal case of a nanoshell in vacuum, described by
stress-free boundary conditions at both interfaces, we have 
$\alpha_c=\alpha_m=\eta_m=\eta_c=0$ and $\chi_c=\chi_m=4\beta_s^2$.
For a thin nanoshell, $1-\kappa\ll 1$, we then easily
recover the known expression for the fundamental mode \cite{love-elast}
\begin{equation}
\xi_0 = 2 \beta_s \sqrt{3 - 4\beta_s^2}.
\label{freeeigen}
\end{equation}
The eigenvalue is, of course, purely real since no energy leaks
through the interface. 

In the realistic case of a nanoshell in a medium, the role of
dielectric core is diminished. The laser pulse causes faster expansion
of the metal shell than of dielectric core. Indeed, the
heating of Au lattice is due to the  cooling of electron gas that occurs
during several ps, while the heating of dielectric core due to heat
transfer from metal to  dielectric takes place on a longer time
scale\cite{vallee-jpcb01}. Because a larger metal thermal expansion
coefficient as compared to 
that of core dielectric, the shell expands to a greater extent than
the dielectric core. As a result, in the new equilibrium, the core and
the shell are, in fact, no longer in contact, so the boundary
conditions at he core/shell interface should be 
be taken as stress-free. This can be accomplished by setting
$\eta_c=0$. For a thin nanoshell, $1- \kappa\ll 1$,
Eqs. (\ref{breathmodes}) can be then reduced to 
%
\begin{equation}
\frac{\chi_c}{1-\kappa}
\Biggl( 
\chi_m -\chi_c + \frac{\alpha_m \eta_m \xi^2}{\alpha_m - i \xi} 
\Biggr)
= 
\Biggl( 
\chi_m + \frac{\alpha_m \eta_m \xi^2}{\alpha_m - i \xi}
\Biggr) \xi_0^2 
- \chi_c \xi^2 
.
\label{simple}
\end{equation}
Typically, the density of the metal shell is much larger than that of
the surrounding dielectric medium, i.e., the parameter $\eta_m$ is small. For
$\eta_m\ll 1$, using $\chi_m-\chi_c = -4 \eta_m \alpha_m^2\beta_m^2$
and $\chi_m/\chi_c = 1 - \eta_m \beta_m^2$, we obtain
\begin{equation}
x^2 - 1  = 
\frac{\alpha_m \eta_m}{\xi_0 (1 - \kappa)} 
\Biggl[ 
\frac{4 \alpha_m \beta_m^2}{\xi_0} - \frac{x^2}{\alpha_m/\xi_0 - i x}
\Biggr],
\label{simple2}
\end{equation}
where $x = \xi/\xi_0$. Note that although both $1-\kappa$ and $\eta_m$
are small, their ratio can be arbitrary. It is now easy to see
that there are two regimes governed by the parameter
\begin{equation}
\lambda=
\frac{\alpha_m \eta_m}{\xi_0 (1 - \kappa)}.
\label{lambda}
\end{equation}
For a very thin nanoshell, $\lambda\gg 1$, the lowest eigenvalue is given by
\begin{equation}
\xi\simeq 2\alpha_m\beta_m\Bigl( \sqrt{1-\beta_m^2} -i \beta_m\Bigr).
\label{lambda-large}
\end{equation}
In this regime, the energy and damping are comptetely determined by
the surrounding medium and are independent of nanoshell aspect ratio. Note
that if the medium transverse sound speed is zero (e. g., in
water), then $\beta_m=c_T^{(m)}/c_L^{(m)}=0$ and both the energy and
damping rate vanish. In the second regime, corresponding to $\lambda\ll 1$,
the spectrum can be obtained as 
%
\begin{eqnarray}
\xi \simeq
\xi_0 - \frac{\lambda}{2}
\Biggl[ \frac{\alpha_m+i\xi_0}{(\alpha_m/\xi_0)^2 + 1}- 4 \alpha_m\beta_m^2
\Biggr], 
\label{lambda-small}
\end{eqnarray}
Here the real part, in a good approximation, is given by
${\rm Re} \xi\simeq \xi_0$, and is independent of medium or aspect
ratio, while the imaginary part, although small
(${\rm Im} \xi\ll {\rm Re} \xi$), depends on both. 
Putting all together, we obtain in this regime
%
\begin{eqnarray}
\omega & \simeq & \frac{c_L^{(s)}}{R_2}\, 2 \beta_s \sqrt{3 - 4\beta_s^2}
\nonumber\\
\gamma & \simeq & \frac{c_L^{(m)}}{d}\,
\frac{2\eta_m \beta_s^2 (3-4\beta_s^2)}{\alpha_m^2 +
  4\beta_s^2(3-4\beta_s^2)},  
\label{lambda-relevant}
\end{eqnarray}
where $d$ is the shell thickness. Thus, for thin nanoshells, the damping
rate is determined by the shell thickness rather than the overall
size.

\section{Numerical results and discussion}
\label{sect:num}  

Here we present the results of our numerical calculations of
vibrational mode spectrum for Au nanoshells in water. The following
material parameters were used: longitudinal sound speed in gold
$c_L^{(s)} = 3240$ m/s, transverse sound
speed $c_T^{(s)} = 1200$ m/s, the density of gold $\rho^{(s)} = 19700$
kg/m$^3$; corresponding values for water are $c_L^{(m)}= 1490$ m/s,
$c_T^{(m)} = 0$, and $\rho^{(m)} = 1000$ kg/m$^3$.

\begin{figure}[t]
\begin{center}
 \includegraphics[width=2.5in]{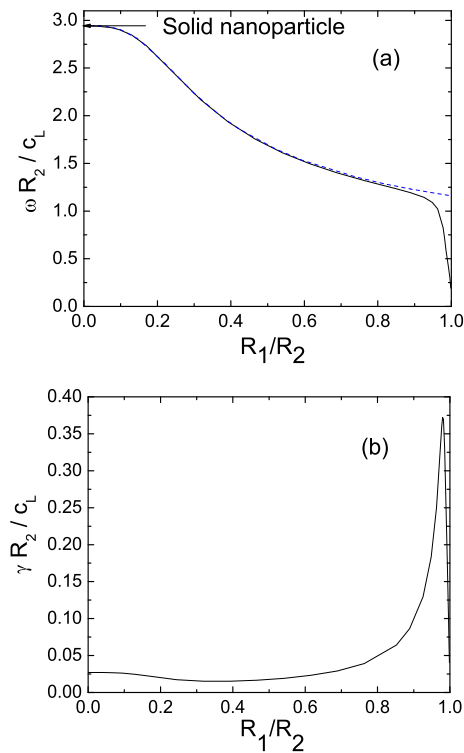}
\end{center}
\caption{\label{fig:spectrum-sym}Spectrum for fundamental breathing
  radial mode in gold nanoshell versus its aspect ratio $R_1/R_2$. 
(a) Solid line: eigenfrequency versus $R_1/R_2$ in the model
with free inner boundary and ideal contact between outer shell and
matrix. Dashed line: eigenfrequency in the model with free boundaries.
(b) Solid line: normalized damping rate versus aspect ratio.
}
 \end{figure}

\begin{figure}[bht]
\centering
 \includegraphics[width=2.5in]{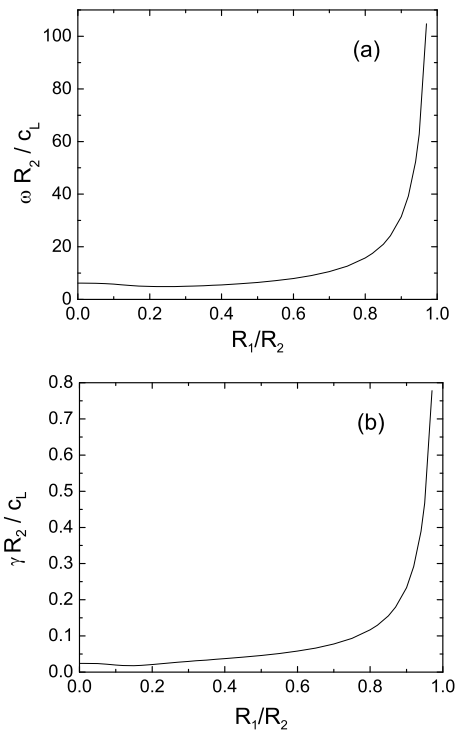}
\caption{\label{fig:spectrum-antisym}Spectrum for $n=1$ radial
  mode in gold nanoshell versus its aspect ratio $R_1/R_2$. 
(a) Eigenfrequency calculated with free inner boundary and ideal
contact between outer shell and medium.  
(b) Normalized damping rate versus aspect ratio.
} 
\end{figure}

In Fig. \ref{fig:spectrum-sym} we show the energy and damping rate for
fundamental ($n=0$) breathing mode versus aspect ratio
$R_1/R_2$. After the initial drop by factor of two, the
frequency changes weakly for thin nanoshells in the aspect ratio range
0.6 - 0.9. In the same range, the damping rate increases by factor of
four. A sharp change in behavior for very thin nanoshells with
$R_1/R_2=0.95$ indicates the transition to overdamped 
regime when stored energy is instantly transferred to the surrounding
medium. Note that for water
($c_T=0$) both energy and damping rate vanish in the thin shell
limit. In contrast, for $n=1$ mode, no such transition takes place,
and both energy and damping rate increase with aspect ratio, as shown
in Fig.\ \ref{fig:spectrum-antisym}. 

%
\begin{figure}[thb]
\centering
 \includegraphics[width=2.5in]{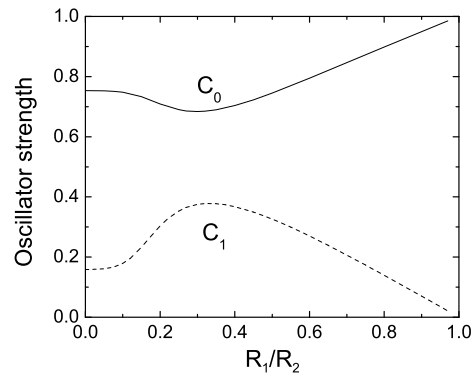}
\caption{\label{fig:oscillator}
Oscillator strengths for symmetric (solid line) and antisymmetric
(dashed line) breathing modes 
of nanoshell versus nanoshell aspect ratio $R_1$/$R_2$.
At $R_1$ = 0 oscillator strength coincide with that for solid nanoparticle.
} 
\end{figure}

Let as now turn to oscillator strengths of fundamental and
excited modes. The radial displacement $\delta r(t)=r(t)-r_0$ of a pont
parcticle inside the shell from its equilibrium position $r_0$ is
given by a general expansion 
$\delta r(t)=\sum_n b_nv_n(r)e^{-i\omega_nt}$, where
$v_n(r)=u_n(r)/\bigl(\int u_n^2dV\bigr)^{1/2}$ is normalized eigenfunction
of the mode with frequency $\omega_n$. To find coefficients $b_n$
that determine the relative contribution of $n$th mode, one has to
specify the initial condition. We assume that the laser spot is much
larger than the nanoshell overall size so that the intial rapid
expansion is homogeneous, $\delta r(0)\propto r$. Then the expression for
the $n$th mode oscillator strength has the form 
\begin{equation}
C_{n}= \frac{b_n}{\bigl(\sum_n b_n^2\bigr)^{1/2}}= 
\frac{ \langle r^2\rangle^{-1/2}\int r \, u_n dV}
{ V^{1/2} \Bigl[\int u_n^2 dV\Bigr]^{1/2}},
\label{oscill}
\end{equation}
where
$\langle r^2\rangle= V^{-1} \int
r^2dV=R_2^2\frac{3(1-\kappa^5)}{5(1-\kappa^3)}$ 
and $V$ is the shell volume. In Fig. \ref{fig:oscillator}, we
show calculated oscillator strengths for $n=0$ and $n=1$ modes versus
aspect ratio. In contrast to solid particles, where the relative
strengths of two modes is constant, $C_1/C_0=1/4$, here $C_1$ vanishes
in the $\kappa=1$ limit. 
For $1-\kappa\ll 1$, 
the aspect ratio dependence of $C_0$ can be found analytically as
\begin{equation}
C_{0} = (1+\kappa)/2.
\label{c0}
\end{equation}
In the $\kappa=1$ limit,  $C_0$ reaches it maximal value, $C_0=1$,
i.e., the fundamental mode carries the entire
oscillator strength. As a result, in nanoshells, excitation of the
fundamental mode should results in a greater amplitude of oscillations
as compared to solid particles, while the $n=1$ mode should be
considerably weaker.

This work was supported by NSF under Grant No. DMR-0305557, by NIH
under Grant No. 2 S06 GM008047-33, and by ARO under Grant No.
DAAD19-01-2-0014.

\end{document}